\documentclass[english]{iopart}
\usepackage[T1]{fontenc}
\usepackage[latin9]{inputenc}
\usepackage{float}
\usepackage{amssymb}
\usepackage{graphicx}
\usepackage{esint}

\makeatletter

\newcommand{\binom}[2]{{#1 \choose #2}}

\usepackage{iopams}
\usepackage{setstack}

\makeatother

\usepackage{babel}
\begin{document}

\title{A geometric wave function for few interacting bosons in a harmonic trap}

\author{B. Wilson\textsuperscript{1}, A. Foerster\textsuperscript{1,2},
C. C. N. Kuhn\textsuperscript{2,3}, I. Roditi\textsuperscript{4}
and D. Rubeni\textsuperscript{2}}

\address{\textsuperscript{1}Department of Theoretical Physics, Research School
of Physics and Engineering, Australian National University, Canberra
ACT 0200, Australia}

\address{\textsuperscript{2}Instituto de F\'{\i}sica da UFRGS, Porto Alegre, RS
- Brazil}

\address{\textsuperscript{3}Quantum Sensors Lab, Department of Quantum Science,
Australian National University, Canberra 0200, Australia}

\address{\textsuperscript{4}Centro Brasileiro de Pesquisas F\'{\i}sicas, Rua Dr.
Xavier Sigaud 150, 22290-180 Rio de Janeiro, Rio de Janeiro, Brazil}
\begin{abstract}
We establish a new geometric wave function that combined with a variational
principle efficiently describes a system of bosons interacting in
a one-dimensional trap. By means of a a combination of the exact wave
function solution for contact interactions and the asymptotic behaviour
of the harmonic potential solution we obtain the ground state energy,
probability density and profiles of a few boson system in a harmonic
trap. We are able to access all regimes,
ranging from the strongly attractive to the strongly repulsive one
with an original and simple formulation.
\end{abstract}


\maketitle

\section{Introduction}

Impressive developments in the preparation and control of traps by the application of dipole lasers \cite{Boyer, Chin}, along with cooling techniques, unveiled the richness of  phenomena occurring in ultracold physics and strengthened our insight into the physical properties of quantum matter. Several aspects of quantum many-body physics that where known only on theoretical grounds are now finding their way into the laboratory, in particular an increasing set of exactly solved models \cite{guan,batch,he,Batchelor}. At present, following this direction a renewed interest is emerging on quantum few body systems. The model explored in our work, that of a set of bosonic atoms trapped in a one-dimensional (1D) harmonic potential and interacting through a contact potential, at arbitrary strength for both the attractive and repulsive regimes, emerges as a particularly fundamental one. Few-body quantum systems can be placed among some of the underlying building blocks of matter. Notwithstanding their simplicity the study of such systems has repeatedly been challenging \cite{blume}. In the case of the harmonically trapped system, even for just three particles, there is no analytical solution of the Schr\"odinger equation. Better means of understanding the physics of such systems are bound to become crucial as, currently, there are a number of techniques which may be applied to trap bosonic systems in a quasi-1D regime \cite{Rigol} and it is reasonable to expect that, as has been the case for fermions \cite{Jochim1,Jochim2}, very soon an ensemble of few bosons will be observed.

The kind of setup able to deal with 1D cold gases, such as optical lattices using lasers with periodic intensities \cite{Kinoshita, Paredes, Bloch} paved the way to physically realize exactly solvable models establishing new levels of interaction between theory and experiments. In optical lattices, the level of control is such that by simply changing their spatial configuration it is possible to tune the dimensionality from 1D to 3D \cite{guan}. Two paradigmatic important achievements in controlling experimental parameters were the observation of a quantum phase transition to the highly-correlated Mott insulator state from the superfluid state for a gas of {$^{87}$Rb} atoms \cite{Greiner1} as well as the tuning between a Bardeen-Cooper-Schrieffer (BCS) superfluid and a Bose-Einstein condensate (BEC) attained when cooling a fermonic gas of  {$^{40}$K} to a quantum degenerate state  \cite{Greiner2}. The range of subjects that falls into this category is wide, and includes mesoscopic systems such as quantum dots, molecular clusters as well as nano-physics. In light of these prospects many interesting purely theoretical breakthroughs are under scrutiny and revealing their usefulness for the new data. Some few examples are the Lieb-Liniger solution for the interacting Bose gas, the Tonks-Girardeau gas and the Super-Tonks gas \cite{Kinoshita, Paredes, Haller, Lieb, Girardeau, Astra}.

In this paper we present a wave function, inspired by the exact solution for a system of bosons interacting through a contact potential and apply it  to the case where this system is confined in a harmonic trap. Although in the presence of a harmonic trap this system is not exactly solved we assume that, in a certain region,  in the low density case, the scattering of the trapped constituents is dominantly non-diffractive \cite{Lamacraft, Sutherland} and hence the exact solution provides an optimal description. In this way, we propose a {\it geometrical} variational wave function in the sense that where the contact interaction is dominant, this wave function is the exact one for a non-diffractive regime, while in the region where the harmonic trap is dominant, the quantum system is described by a smoothly decreasing function. This idea is a sharp advance with respect to a one used for just two fermions that already provided interesting results \cite{Rubeni} (see also \cite{chines}). For the ground state of particles in a harmonic trap, an exact analytical solution only exists for the two-body case \cite{Busch} (see also \cite{Calarco}). This solution was explored in an analytical ansatz for a few-boson system wave function \cite{Schmelcher} based on a number of assumptions, respecting the analytically known limits of zero and infinite repulsion. Ours, on the other hand, is based on the complete knowledge of the exact solution for the Bose and Fermi gas interacting via a delta-function term (contact potential) \cite{Lieb,Yang, Gaudin, Takahashi}, and through a variational calculation we are able to handle from the repulsive to the attractive regime. In the following, we present our construction, and results, for two and three bodies in both the attractive and repulsive regimes. We notice that its geometrical nature with only two regions makes the extension to a different number of particles attainable, even if, as expected, it may involve a more complex set of numerical calculations.

\section{The model}

In the ultracold region the study of trapped atoms is slightly simplified since, in this case, the de Broglie wavelength is large enough to allow the description of a complex interaction by a simple contact potential that can be modelled as a delta function.

Let us then consider a system of $N$  interacting bosons with mass $m$, in an axially symmetric harmonic trap with angular frequency $\omega$. Such a system is described by the following Hamiltonian in absolute coordinates  

\begin{equation}
\!\!\! H\!=\!\sum_{i=1}^{N}\left(-\frac{\hbar^{2}}{2m}\frac{\partial^{2}}{\partial x_{i}^{2}}+c\sum_{j<i}\delta\left(x_{i}-x_{j}\right)+\frac{1}{2}m\omega^{2}x_{i}^{2}\right)\!,\label{hamiltonian1}
\end{equation}
where $c$ is the interaction strength, repulsive for $c>0$ and attractive for $c<0$.  In spite of its simplicity, the harmonic potential term, $\frac{1}{2}m\omega^{2}x_{i}^{2}$, prevents the exact solvability of the above Hamiltonian. If we consider just the interaction Hamiltonian,

\begin{equation}
H_{I}=\sum_{i}\left(-\frac{\hbar^{2}}{2m}\frac{\partial^{2}}{\partial x_{i}^{2}}+c\sum_{j<i}\delta\left(x_{i}-x_{j}\right)\right),\label{hamiltonian2}
\end{equation}
we have the description of a system which is exactly solvable by means of the Bethe ansatz. In the region $\chi$ such that $x_{1}<x_{2}<...< x_{N}$, the solution is given by the following wave function \cite{Takahashi}

\[
\psi_{\chi}\left(x_{1},x_{2},\dots,x_{N}\right)=\sum_{P}A(P)\exp(i(\mathrm{k}_{P1}x_{1}+\dots+\mathrm{k}_{PN}x_{N}))
\]

\begin{equation}
A(P)=C\epsilon(P)\prod_{j<l}(\mathrm{k}_{Pj}-\mathrm{k}_{Pl}+ic)\label{taka}
\end{equation}
where the sum is over all permutations of the quasi-momenta, $\mathrm{k}_i, i=1,... N$, and $\epsilon$ is the Levi-Civita symbol (1 or -1 for, respectively, even or odd permutations).  The complete Bethe ansatz wave function $\psi_{B}$  for all regions can be determined by the full symmetry of the wave function (see \cite{Takahashi} for more details). The above has been a pivotal solution for interacting gases in 1D, and as we shall see a central part of our ansatz.

\section{A geometrical ansatz}

\begin{figure}[H]
\begin{centering}
\includegraphics[width=6cm]{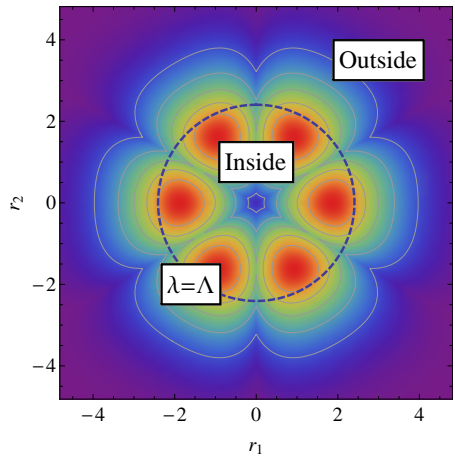}
\par\end{centering}

\caption{Schematic representation of the probability density  $|\Psi|^{2}$  for the three bosons case.  The variational parameter $\Lambda$ determines the boundary between two regions: inside (Bethe ansatz) and outside (asymptotic harmonic oscillator). The colors range from purple to red indicating respectively lower values and higher values of $|\Psi|^{2}$. \label{fig:AnsatzThreeBoson}}
\end{figure}

Due to the symmetry of our system it is convenient to move to Jacobi coordinates \cite{Werner}, which allows us to remove the centre of mass coordinate and re-express the remaining coordinates as a set of relative coordinates.  The general coordinate transformation from Cartesian to Jacobi coordinates is

\begin{eqnarray}
R & = & \frac{1}{N}\sum_{i=1}^{N}{x_{i}}\\
r_{1} & = & x_{2}-x_{1}\\
r_{j} & = & \binom{j+1}{j-1}^{-1/2}\left(jx_{j+1}-\sum_{i=1}^{j}{x_{i}}\right)
\end{eqnarray}
with $j = 1, ... N-1$. We should note that the unconventional factors are chosen so that the effective mass of each coordinate is the same. We will now refer to $\vec{r}=\{r_1,r_2,\dots,r_{N-1}\}$ as the relative coordinates and $R$ as the centre of mass coordinate. With this choice the Hamiltonian can be written as

\begin{eqnarray}
H & = & -\frac{\hbar^{2}}{2M}\frac{\partial^{2}}{\partial R^{2}}+\frac{1}{2}M\omega^{2}R^{2}\nonumber \\
 &  & +\sum_{i=1}^{N-1}\left(-\frac{\hbar^{2}}{2\mu}\frac{\partial^{2}}{\partial r_{i}^{2}}+\frac{1}{2}\mu\omega^{2}r_{i}^{2}\right)+c\sum_{j}\delta\left(d_{j}(\vec{r})\right)
\end{eqnarray}
where $M=N m$, $\mu=\frac{1}{2}m$, and $d_j$ are the locations of the delta function interactions in the relative coordinate system. These are hypersurfaces which extend radially from the origin. The above Hamiltonian is separable, meaning that it is possible to solve the Schr\"odinger equation by separating the center of mass and the relative coordinates. The relative motion part of the Hamiltonian has an approximate radial symmetry, so we make an additional change of coordinates to hyperspherical coordinates. In these coordinates we shall denote the corresponding radial component $\lambda$ and the angular part $\vec{\theta}=\{\theta_1,\theta_2,...\theta_{N-2}\}$. The relative Hamiltonian now takes the form:

\begin{equation}
\hat{H}_{rel}=-\frac{\hbar^{2}}{2\mu}\nabla^{2}+\frac{1}{2}\mu\omega^{2}\lambda^{2}+c\sum_{j}\delta\left(d_{j}(\vec{\theta})\right).
\end{equation}
Looking at the above Hamiltonian, we observe that for small $\lambda$, the Hamiltonian is approximately the Hamiltonian of $N$ bosons with a delta function potential, the one solved by the Bethe ansatz. However, for sufficiently large $\lambda$ the Hamiltonian behaviour is dominated by that of a harmonic oscillator, and in this limit we expect a Gaussian decay of the wave function. We therefore make the following ansatz for the wave function 

\begin{equation}
\Psi(\lambda,\vec{\theta})=\left\{ \begin{array}{cc}
\psi_{B}\left(\vec{\kappa},\lambda,\vec{\theta}\right) & \lambda<\Lambda\\
A(\vec{\theta})\exp\left(-\alpha(\vec{\theta})(\lambda^{2}-\Lambda^{2})\right) & \lambda>\Lambda
\end{array}\right..
\end{equation}
Above $\psi_{B}$ is the complete Bethe ansatz wave function in the relative coordinates system  after the change to the hyperspherical coordinates, $\Lambda$ is a parameter which determines the boundary between the inside Bethe ansatz and the outside harmonic oscillator regions, $\alpha(\vec{\theta})$ is the Gaussian decay parameter which is used to match the derivative at the boundary of the two regions, and $\vec{\kappa} = \{\kappa_1,... \kappa_{\lfloor N/2\rfloor}\}$ are parameters which originated from the Bethe ansatz wave function. They are similar to those in Eq.(\ref{taka}) but considered as parameters instead of quasi-momenta, and the number of parameters is lowered as we are considering the string hypothesis \cite{Lieb,Yang, Gaudin,Takahashi}. Both $\Lambda$ and $\vec{\kappa}$ will be used as variational parameters in the  minimization procedure used to obtain the ground state energy. In Fig. \ref {fig:AnsatzThreeBoson} we illustrate in a schematic way the probability density obtained from our ansatz for the three bosons case in the repulsive regime. There we can see how the two regions are defined, it is also possible to visualize the hexagonal structure of lower and high probability regions that reflects the symmetry of the pairwise contact interaction in the case of three particles.

Continuity of the wave function and its derivative at the boundary requires

\begin{eqnarray}
A(\vec{\theta}) & = & \psi_{B}(\vec{\kappa},\Lambda,\vec{\theta})\\
\alpha(\vec{\theta}) & = & -\frac{1}{2\Lambda}\frac{1}{\psi_{B}(\vec{\kappa},\Lambda,\vec{\theta})}\left.\frac{\partial\psi_{B}(\vec{\kappa},\lambda,\vec{\theta})}{\partial\lambda}\right|_{\lambda=\Lambda}
\end{eqnarray}
where for two particles, this condition on $\alpha$ is equivalent to the  Bethe Ansatz-type  equations for the boundary condition we have here.

\section{Variational approach}

\begin{figure}[H]
\begin{centering}
\includegraphics[width=8cm]{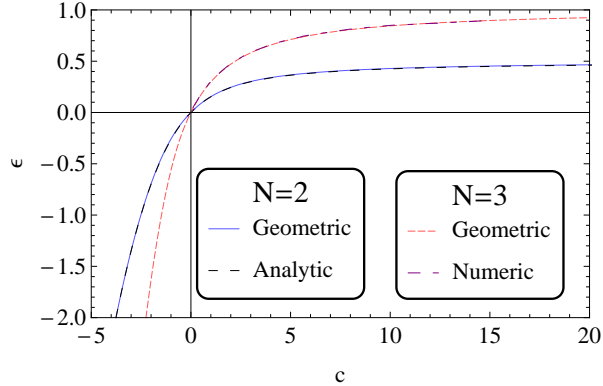}
\par\end{centering}

\caption{Ground state energies $\epsilon=\frac{En}{N\hbar\omega}-\frac{1}{2}$ as a function of the interaction strength $c$ for different number of bosons $N$. The case $N=2$ matches the analytic result \cite{Busch,Calarco} well in all regimes, and the $N=3$ case matches published results in the repulsive regime \cite{Schmelcher}. \label{fig:epsilonvsc}}
\end{figure}

Next we perform the variational method to minimise the energy for this ansatz to find an approximation for the ground state energy. This provides an upper bound on the ground state energy of the actual system. 

\begin{eqnarray}
\left\langle \Psi\right|\hat{H}_{rel}\left|\Psi\right\rangle  & = & \int\mathrm{d}\lambda\mathrm{d}\Omega\,\lambda^{N-1}\Psi^{*}(\lambda,\vec{\theta})\hat{H}_{rel}\Psi(\lambda,\vec{\theta})\\
\,\,\,\,\,\,\,\,\,\,\,\,\,\,\left\langle \Psi\right|\left.\Psi\right\rangle  & = & \int\mathrm{d}\lambda\mathrm{d}\Omega\,\lambda^{N-1}\Psi^{*}(\lambda,\vec{\theta})\Psi(\lambda,\vec{\theta})
\end{eqnarray}
where $\Omega$ is the solid angle. Note that by construction, the delta function in the internal region cancels with the discontinuity of the derivative of the wave function contribution, however this exact cancelation is not exact in the external region. We can find the contribution from the discontinuous derivative along the delta functions by integrating the kinetic energy term in the Hamiltonian by parts

\begin{eqnarray}
\int_{0}^{\infty}\mathrm{d}\lambda\int\mathrm{d}^{N-2}\Omega\,\lambda^{N-1}\Psi^{*}(\lambda,\vec{\theta})\nabla^{2}\Psi(\lambda,\vec{\theta})=\nonumber \\
\sum_{j}\int_{0}^{\infty}\mathrm{d}\lambda\int_{d_{j}(\vec{\theta})=0}\mathrm{d}^{N-3}\Omega\,\lambda^{N-1}\Psi^{*}(\lambda,\vec{\theta})\nabla\Psi(\lambda,\vec{\theta})\cdot\hat{\mathbf{n}}
\end{eqnarray}
where $\hat{\mathbf{n}}$ is the surface normal. The radial integral over $\lambda$ can be performed analytically. However the angular integral must be evaluated numerically. The ground state for the wave function is then found by minimising the energy with respect to the variational parameters, $\Lambda$ and $\vec{\kappa}$.

\begin{eqnarray}
\left.\frac{\partial}{\partial\Lambda}\frac{\left\langle \Psi\right|\hat{H}_{rel}\left|\Psi\right\rangle }{\left\langle \Psi\right|\left.\Psi\right\rangle }\right|_{\Lambda=\Lambda^{*}} & = & 0\\
\left.\frac{\partial}{\partial\vec{\kappa}}\frac{\left\langle \Psi\right|\hat{H}_{rel}\left|\Psi\right\rangle }{\left\langle \Psi\right|\left.\Psi\right\rangle }\right|_{\vec{\kappa}=\vec{\kappa}^{*}} & = & 0
\end{eqnarray}
In this way we determine the ground state energy of a few bosons system in a harmonic trap as a function of the coupling $c$ via the variational principle, where the trial wave function is constructed by combining  the Bethe ansatz and the asymptotical behaviour of the  harmonic oscillator . This result is depicted in  Fig. \ref{fig:epsilonvsc} and a very good agreement is found with the analytical solution for $N=2$ particles \cite{Busch,Calarco} and existing results in the repulsive regime for $N=3$ \cite{Schmelcher}.

\begin{figure}[H]
\begin{centering}
\includegraphics[width=9cm]{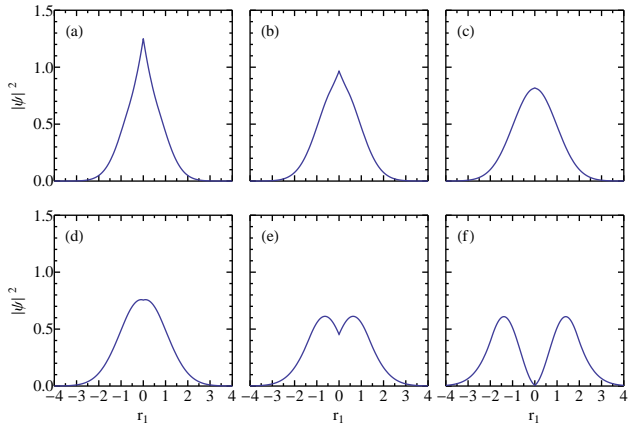}
\par\end{centering}

\caption{Probability density for the two bosons case for different values of the coupling constant (a) $c=-1$, (b) $c=-0.5$, (c) $c=-0.1$, (d) $c=0.1$, (e) $c=1$ and (f) $c=20$. \label{TwoBosonsWavefunction}}
\end{figure}

\begin{figure}[H]
\begin{centering}
\includegraphics[width=9cm]{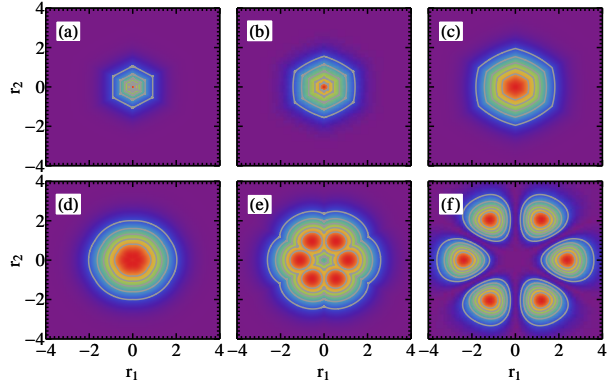}
\par\end{centering}

\caption{Probability density on the frame of $r_1$ and $r_2$ Jacobi coordinates  for the three bosons case  for different values of the coupling constant (a) $c=-1$, (b) $c=-0.5$, (c) $c=-0.1$, (d) $c=0.1$, (e) $c=1$ and (f) $c=20$. The colors range from purple to red indicating respectively lower values and higher values of the probability density \label{ThreeBosonsDensityr1r2}}
\end{figure}

In Figs \ref{TwoBosonsWavefunction} and \ref{ThreeBosonsDensityr1r2} we present the  probability density of two and three bosons , respectively, in the ground state for different couplings, ranging from the attractive to the repulsive regimes: $c=(-1,-0.5, -0.1, 0.1, 1, 20)$.

For two particles the probability density of the relative motion in the attractive case exhibits a peak at $r_1=0$ which increases and gets thinner for higher $|c|$ values, while for the repulsive case a cusp emerges at $r_1=0$ which goes to zero by increasing $c$. Similarly, for three particles with attractive interaction a more localized peak in the probability density is observed by increasing $|c|$, whereas  along the mirror planes \cite{Sutherland} (the points where $x_i-x_j=0$ 
for $i \neq  j$) it can be observed that, in the repulsive case, the probability density reduces when $c$ increases. 
We can clearly see that our ansatz captures the most relevant aspects regarding the physical properties of the studied system: in the repulsive case,  the tendency of the particles to repel and thus to stay away from each other when the  interaction strength $c$ is increased; while in the attractive case the gregarious tendency of the particles is increased for higher $|c|$ values.

\begin{figure}[H]
\begin{centering}
\includegraphics[width=9cm]{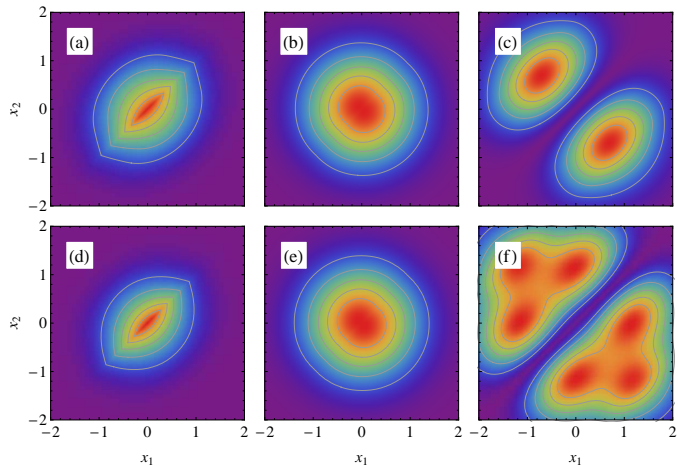}
\par\end{centering}

\caption{Pair correlation function $\rho_2(x_1,x_2)$ for the two bosons case (upper line) for different values of the coupling constant (a) $c=-5$, (b) $c=0.1$, (c) $c=20$ and for the three bosons case (bottom line) for  (d) $c=-5$, (e) $c=0.1$ and (f) $c=20$. The colors range from purple to red indicating respectively lower values and higher values of the two-body density.
\label{TwoAndThreeCorrelations}}
\end{figure}

\begin{figure}[H]
\begin{centering}
\includegraphics[width=8cm]{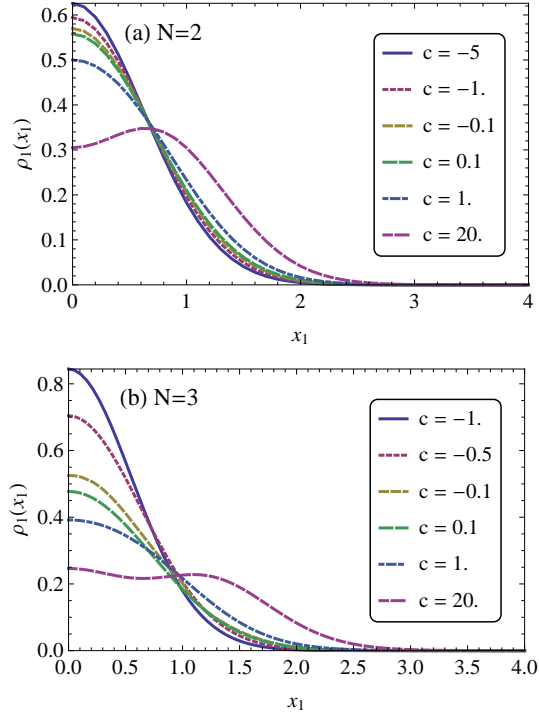}
\par\end{centering}

\caption{Normalized one-body density $\rho_1(x_1)$ as a function of the position $x_1$ for both attractive and repulsive regimes for 
(a) two bosons and  (b) three bosons. 
\label{TwoAndThreeBosonsDensityx1}}
\end{figure}

In Figs \ref{TwoAndThreeCorrelations} and  \ref{TwoAndThreeBosonsDensityx1}  we plot the pair correlation function
$\rho_2(x_1,x_2)$  \cite{Sutherland} and the normalized one-body  density of bosons $\rho_1(x_1)$  obtained by integrating over all coordinates except two, and one, respectively. For $N=2$ $\rho_2(x_1,x_2)$ gives the full probability density as a function of the particle positions  $x_i$, while for  $N=3$ it provides a two-body probability density.

We observe that in the attractive case the pair correlations collapse towards the ground state of a single particle with mass $M$ while in the repulsive case the particles tend to repel and begin to distribute themselves along the trap. In the strong repulsive regime the densities split into different broad lobes  separated by the mirror plane $x_1-x_2=0$ as expected due to the repulsion between the particles. 
These results provide a direct scheme for comparison with density profiles that could be obtained experimentally. 

\section{Conclusion}

We have presented in this paper a geometric wave function that has been set up using the exact solution of the boson gas interacting via a delta potential combined with a smoothly decreasing function. This combination involves only two regions: one where the system is described by the exact solution and another one where the harmonic trap becomes dominant. We emphasize that this geometric wave function, which we use as a variational ansatz, naturally captures the essential physics of the problem, and allowed us to obtain an impressive accord with the numeric benchmark in the case of three bosons, and, in the case of two bosons, with the exact result. Remarkably, it is also valid for both the attractive and repulsive regimes, which is certainly of value if one tries to apply it to the study of excited states, where important physical information could be extracted, such as the nature of the super-Tonks Girardeau gas, for which an excited phase with highly attractive interactions is present. With some modifications, our proposal can be adapted to different scenarios, such as: other trap geometries; fermionic system; mixture systems, composed of bosons and fermions.

\ack{}{The authors acknowledge financial support from CAPES (Proc. 10126-12-0), CNPq, and FAPERJ.   They also thank I. Brouzos and P. Schmelcher for providing some of their data. A. F. thanks M. T Batchelor, {X.W. Guan} and Y. Levin for many helpful discussions. I.R. thanks I.S. Oliveira for an interesting exchange of views.}

\section*{References}{}

\end{document}